\pdfoutput=1
\documentclass[namedreferences,hyperref,optionalrh]{spr-sola}
\usepackage{graphicx}        % For eps figures, newer & more powerfull
\usepackage{amsmath}
\usepackage{amssymb}
\usepackage{color}           % For color text: \color command
\chardef\us=`\_

\begin{document}

\begin{frontmatter}
\title{Effects of harmonic magnetic field boundary conditions in mean-field solar dynamo}
\author[addressref={aff1},email={valery.pipin@gmail.com}]{\inits{V.V.}\fnm{Valery}~\snm{Pipin}\orcid{0000-0001-9884-1147}}
\address[id=aff1]{Institute solar-terrestrial physics, Irkutsk,
  Russia}

\runningauthor{V.V. Pipin}
\runningtitle{Harmonic boundary conditions in  solar dynamo}
\begin{abstract}
We consider effects of the harmonic magnetic field boundary conditions
at the top of the dynamo domain on the dynamo stability and inside
the solar convection zone. These boundary conditions allow us to
quantify the helical properties of the coronal magnetic field that
stems from the dynamo region. In sewing the tangential component of
the mean electric field we are able to take into account the effect the
diffusive properties of the stellar corona on the dynamo instability.
The model shows that effect of the vacuum boundary conditions can be
restored if we introduce a few orders of magnitude jump of the coronal
magnetic field turbulent diffusion over its typical value at the top
of the dynamo domain. The parameters of this jump define the critical
instability threshold of the $\alpha$ effect in the $\alpha^{2}\Omega$
dynamo.
\end{abstract}
\end{frontmatter}

\section{Introduction}

Observations indicate a nonzero toroidal magnetic field on the photosphere
of the Sun and other stars and an increase in field amplitude
with increasing stellar rotation velocity \citep{Duvall1979,See2016}.
Meanwhile, the standard framework of the mean-field dynamo models
suggests the vacuum boundary conditions for the top boundary \citep{Krause1980}.
This results in insulation of the coronal magnetic activity phenomena
from the dynamo operating in the convective envelope of the star.
In the literature there were a number of attempts to find out the
top boundary conditions for the dynamo that would take into account
a nonzero toroidal magnetic field at the surface (see, e.g., \cite{Moss1992,Kitchatinov2000}).
The most consistent approach is to include the coronal magnetic field
in the dynamo simulations. The mean-field dynamo models of \cite{Elstner2020IAUS}
as well as the results of numerical simulations of \cite{JakBran2021AA,Perri2021ApJ}
showed a possible influence of the stellar corona on the properties
of the large-scale dynamo inside the convective envelope. 

In this paper we employ the idea of \cite{Bonanno2016}(hereafter
B16) to study the effects of the diffusive properties of the coronal magnetic
field on the large-scale dynamo instability inside the star's convective
envelope. The harmonic magnetic field outside of the dynamo region
satisfies the differential equation:
\begin{equation}
\Delta\boldsymbol{B}+k^{2}\boldsymbol{B}=0.\label{eq:b16}
\end{equation}
\cite{Bonanno2016} found a decrease in the dynamo instability
threshold for the harmonic magnetic field boundary conditions. For
the case of the dynamo concentrated at the bottom of the convection
zone, the top condition Eq(\ref{eq:b16}) does not affect the properties
of the dynamo waves. In our paper, we study the model of the so-called
distributed dynamo \citep{Brandenburg2005}. This state-of-the-art
mean-field dynamo model was introduced in our earlier papers\citep{Pipin2011,PK19}.
The dynamo model is based on the idea \cite{Parker1955} of large-scale dynamo waves that are excited by means of differential
rotation and turbulent cyclonic convection. Helioseismology shows
that these dynamo waves propagate from the bottom of the convection zone
to the top \citep{Kosovichev2019,Mandal2024}. In this model, we
put no restriction on the distributions of the dynamo effects inside
the stellar convection zone. The nonlinear dynamo model self-consistently
explains many features of the solar activity, including the butterfly
diagrams of the sunspot activity, evolution of the polar magnetic
field, and the zonal variations of the angular velocity in radius and
latitude. 

Our goal is to connect the mean field dynamo model with solar observations
of the surface toroidal magnetic field using the boundary condition
Eq(\ref{eq:b16}). At the top of the dynamo domain, we require the continuity
of the radial magnetic field and the horizontal electric field. The latter condition provides the condition for the toroidal magnetic field.
The solar observations show that the surface toroidal magnetic field
can be a few Gauss. In what follows, we show that the dynamo model
that satisfies the solar observations requires a jump turbulent diffusivity
in the transition from the dynamo region to the corona. Further study shows
that the magnitude of this jump defines the dynamo instability threshold
parameters and some properties of the dynamo waves such as the dynamo
period and the time-latitude evolution of the magnetic activity. Our
plan is as follows. The Section 2 formulates the mean field dynamo
model and the boundary conditions. Section 3 considers the results
for the eigenvalue dynamo instability problem and the numerical solutions of nonlinear dynamo models. The final Section 4 discusses the possible applications of our findings to the solar/stellar coronal activity
phenomena.

\section{Dynamo model and boundary conditions}

\subsection{The model}

\begin{figure*}
 \includegraphics[width=0.95\textwidth]{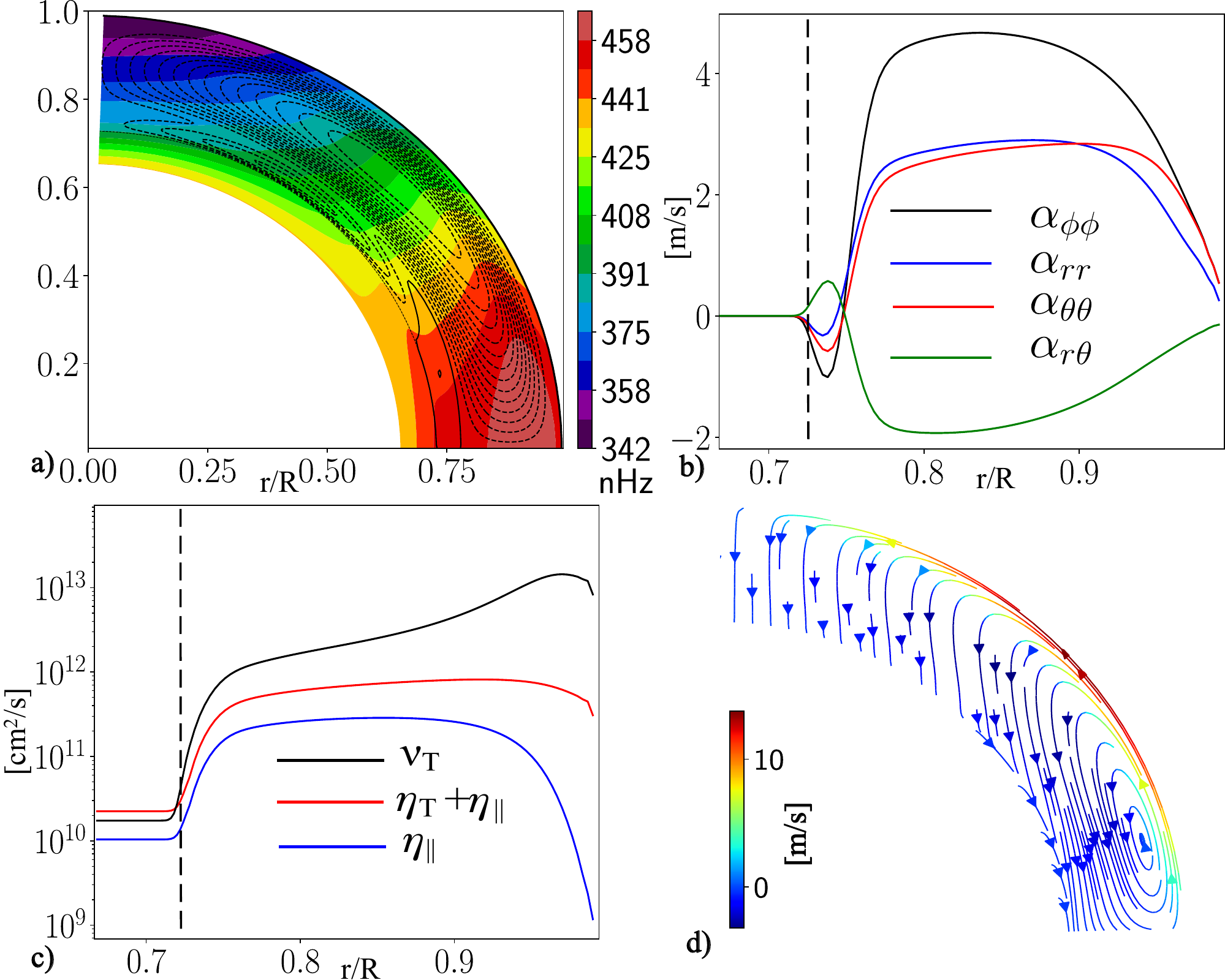} \caption{ a) The meridional circulation (streamlines) and the angular velocity
distributions; the magnitude of circulation velocity is of 13 m/s
on the surface at the latitude of 45$^{\circ}$; b) the $\alpha$-effect
tensor distributions at the latitude of 45$^{\circ}$, the dash line
shows the convection zone boundary; b) radial dependencies of the
total, $\eta_{T}+\eta_{||}$, and the rotational induced part, $\eta_{||}$,
of the eddy magnetic diffusivity, the eddy viscosity profile, $\nu_{T}$;
d) streamlines of the effective drift velocity of large-scale magnetic
field due to the turbulent pumping and the meridional circulation.
We use \textsc{numpy/scipy} \citep{harris2020array,2020SciPyNMeth}
together with \textsc{matplotlib \citep{Hunter2007} }and\textsc{
pyvista \citep{sullivan2019pyvista} }for post-processing and
visualization.}\label{fig1}
\end{figure*}

We model the large-scale dynamo using the mean magnetic field, $\left\langle \boldsymbol{B}\right\rangle $,
induction equation in the turbulent highly conductive media \citep{Krause1980},
\begin{eqnarray}
\partial_{t}\left\langle \boldsymbol{B}\right\rangle  & = & \mathbf{\nabla}\times\left(\mathbf{\boldsymbol{\mathbf{\mathcal{E}}}}+\left\langle \boldsymbol{U}\right\rangle \times\left\langle \boldsymbol{B}\right\rangle \right).\label{eq:dyn}
\end{eqnarray}
Here, $\left\langle \dots\right\rangle $ stands for the ensemble
average of the turbulent field. In this paper, we discuss the axisymmetric
large-scale flow and magnetic field, i.e., 
\begin{eqnarray}
\left\langle \overline{\boldsymbol{B}}\right\rangle  & \equiv & \overline{\boldsymbol{B}},\,\left\langle \overline{\boldsymbol{U}}\right\rangle \equiv\overline{\boldsymbol{U}},\label{eq:b0}
\end{eqnarray}
where the overline $\overline{\dots}$ denotes the azimuthal averaging.
In the model, the large-scale flow $\overline{\boldsymbol{U}}$ includes
the effects of differential rotation and meridional circulation.
To study the large-scale dynamo instability we employ the kinematic
approach, i.e., the flow $\overline{\boldsymbol{U}}$ is given (see
Fig. \ref{fig1}, below). The large-scale magnetic field is decomposed
to the sum of the toroidal and poloidal components: 
\begin{equation}
\overline{\boldsymbol{B}}=B\boldsymbol{e}_{\phi}+\nabla\times\left(\frac{A\boldsymbol{e}_{\phi}}{r\sin\theta}\right)\label{eq:ax}
\end{equation}
where the scalars $B$ and $A$ are the functions of time, $r$ is
the radius, $\theta$ is the colatitude (the polar angle), and $\boldsymbol{e}_{\phi}$
is the unit vector along the azimuth. It is noteworthy that the axisymmetric
magnetic vector potential has two components, 
\begin{eqnarray}
\boldsymbol{\overline{A}} & = & \frac{A}{r\sin\theta}\boldsymbol{e}_{\phi}+\boldsymbol{r}\overline{T},\label{eq:T}\\
\nabla\times\left(\boldsymbol{r}\overline{T}\right) & = & \boldsymbol{e}_{\phi}B.\nonumber 
\end{eqnarray}
The mean electromotive force, $\mathbf{\mathcal{E}}=\left\langle \boldsymbol{u}\times\boldsymbol{b}\right\rangle $
was calculated analytically using the mean-field magnetohydrodynamics
framework \citep{Pipin2008a} (hereafter P08). In the general case,
see \cite{Krause1980}, it reads as 
\begin{equation}
\mathcal{E}_{i}=\left(\alpha_{ij}+\gamma_{ij}\right)\left\langle B\right\rangle _{j}-\eta_{ijk}\nabla_{j}\left\langle B\right\rangle _{k},\label{eq:emf}
\end{equation}
where the first term represents the turbulent generation and turbulent
pumping of the mean magnetic field and the last term represents the
magnetic eddy diffusivity. We will also consider the nonlinear solutions
of the dynamo model. In this case we employ the nonlinear
form of the $\alpha$-effect with the including effect of magnetic
helicity conservation, i.e., 
\begin{eqnarray}
\alpha_{ij} & = & C_{\alpha}\psi_{\alpha}(\beta)\alpha_{ij}^{(H)}+\alpha_{ij}^{(M)}\psi_{\alpha}(\beta)\frac{\left\langle \boldsymbol{a}\cdot\boldsymbol{b}\right\rangle \tau_{c}}{4\pi\overline{\rho}\ell_{c}^{2}},\label{alp2d}
\end{eqnarray}
where the full expressions of the kinetic helicity tensor $\alpha_{ij}^{(H)}$
and the magnetic helicity tensor $\alpha_{ij}^{(M)}$ are given in
P08 and also in \cite{BRetal23}, $\boldsymbol{a}$ and $\boldsymbol{b}=\boldsymbol{\nabla}\times\boldsymbol{a}$
are the fluctuating vector-potential and magnetic field, respectively.
The radial profiles of the $\alpha_{ij}^{(H)}$ and $\alpha_{ij}^{(M)}$
depend on the mean density stratification, the profile of the convective
RMS velocity $u_{c}$ and on the Coriolis number $\Omega^{*}=2\Omega_{0}\tau_{c}$,
where $\Omega_{0}$ is the angular velocity of the star and $\tau_{c}$
is the convective turnover time. The magnetic quenching function $\psi_{\alpha}(\beta)$
depends on the parameter $\mathrm{\beta=\left|\overline{\mathbf{B}}\right|/\sqrt{4\pi\overline{\rho}u_{c}^{2}}}$,
see P08, as well.

To calculate the evolution of $\left\langle \mathbf{a}\cdot\mathbf{b}\right\rangle $
we use the balance equation for the total magnetic helicity, $\left\langle \boldsymbol{a}\cdot\boldsymbol{b}\right\rangle +\overline{\boldsymbol{A}}\cdot\overline{\boldsymbol{B}}$,
(see \citealp{Brandenburg2018,PShKo2025}): 
\begin{equation}
\left(\frac{\partial}{\partial t}+\overline{\boldsymbol{U}}\cdot\boldsymbol{\nabla}\right)\left(\left\langle \boldsymbol{a}\cdot\boldsymbol{b}\right\rangle +\overline{\boldsymbol{A}}\cdot\overline{\boldsymbol{B}}\right)=-\frac{\left\langle \boldsymbol{a}\cdot\boldsymbol{b}\right\rangle }{R_{m}\tau_{c}}+\mathbf{\boldsymbol{\nabla}\cdot}\eta_{\chi}\boldsymbol{\nabla}\left\langle \boldsymbol{a}\cdot\boldsymbol{b}\right\rangle ,\label{eq:helcon}
\end{equation}
where, we use ${\displaystyle 2\eta\left\langle \boldsymbol{b}\cdot\boldsymbol{j}\right\rangle =\frac{\left\langle \boldsymbol{a}\cdot\boldsymbol{b}\right\rangle }{R_{m}\tau_{c}}}$
\citep{Kleeorin1999}. The second term in the RHS defines the
diffusive flux of the small-scale magnetic helicity density, we put
$\eta_{\chi}=\frac{1}{10}\eta_{T}$ \citep{Mitra2010,Kleeorin2022,Subramanian2023ApJ};
$R_{m}$ is the magnetic Reynolds number, we employ $R_{m}=10^{6}$.
For the turbulent diffusivity and turbulent pumping we include effects
of the rotational-induced anisotropy, density stratification, and the
mean-field magnetic buoyancy (for the nonlinear variant of the model).
The interested reader can find their formulation in \cite{BRetal23}.

We calculate the turbulent parameters using the mixing-length approximation
and the profile of the mean entropy. 
\begin{equation}
\mathrm{u_{c}=\frac{\ell_{c}}{2}\sqrt{-\frac{g}{2c_{p}}\frac{\partial\overline{s}}{\partial r}},}\label{eq:uc}
\end{equation}
where $\ell_{c}=\alpha_{\mathrm{MLT}}H_{p}$ is the mixing length,
$\alpha_{\mathrm{MLT}}=1.9$ is the mixing length parameter, and $H_{p}$
is the pressure height scale. The entropy profile is defined by solving
the mean-field heat transport equation (see, e.g., P22) for the rotating
convection zone. It deals with deviations of the mean entropy from
the reference state due to effect of rotation and the heat energy
sink and gain from evolution of the large-scale velocity and magnetic
field. To calculate the reference profiles of the mean thermodynamic parameters,
such as entropy, density, temperature, and convective turnover
time, $\tau_{c}$, we use the MESA model \citep{Paxton2013}.
The Eq.~(\ref{eq:uc}) defines the profiles of the eddy heat conductivity,
$\chi_{T}$, eddy viscosity, $\nu_{T}$, and eddy diffusivity, $\eta_{T}$,
as follows, 
\begin{eqnarray}
\chi_{T} & = & \frac{\ell^{2}}{6}\sqrt{-\frac{g}{2c_{p}}\frac{\partial\overline{s}}{\partial r}},\label{eq:ch}\\
\nu_{T} & = & \mathrm{Pr}_{T}\chi_{T},\label{eq:nu}\\
\eta_{T} & = & \mathrm{Pm_{T}\nu_{T}}.\label{eq:et}
\end{eqnarray}

The large-scale flow distribution is determined by conservation of
the angular momentum and azimuthal vorticity $\overline{\omega}=\left(\mathbf{\nabla}\times\mathbf{\overline{U}}^{(m)}\right)$,
\citep{PK19}. The model shows agreement of the angular velocity
profile with helioseismology results for $\mathrm{Pr}_{T}=3/4$. The
dynamo models with the standard boundary conditions, i.e., the superconductor
at the bottom of the convection zone, $r_{b}=0.728R$, and the vacuum
boundary condition at the top $r_{e}=0.99R$, shows cycle period of
$22$ years when $\mathrm{Pm}_{T}=10$ and the dynamo threshold $C_{\alpha}=0.04$.
It should be noted that the nonlinear dynamo model includes the overshoot
layer down till $r_{i}=0.65R$ with the exponential damping of the
turbulent effects, see \cite{PK19}. To study the dynamo instability
we restrict the dynamo domain in radius by the convective envelope
from $r_{b}=0.728R$ to $r_{e}=0.99R$. 

Figure \ref{fig1} shows the large-scale flow profiles, the
hydrodynamic $\alpha$-effect, the diffusivity profiles, and the effective drift velocity of the toroidal magnetic field. The latter consists of the sum of turbulent pumping an meridional circulation. We note
the inverse sign of the $\alpha$-effect tensor components and the
rotational quenching of the turbulent diffusivity profile toward the
bottom of the convection zone (marked by the dashed line).

\subsection{The outer boundary conditions}

In the solar dynamo models, it is common to employ the vacuum (potential
field) boundary conditions at the top \cite{Krause1980}. Therefore,
in this case, we have $B=0$, at the top. In this case the radial
component of the vector-potential is zero, $\hat{\boldsymbol{r}}\cdot\overline{\boldsymbol{A}}=0$,
or $\overline{T}=0$ at the top boundary, see Eq(\ref{eq:T})). Moreover,
we have $\nabla\cdot\overline{\mathbf{A}}=0$ at the top boundary.
The stellar magnetic activity observations show non-zero the axisymmetric
toroidal magnetic field at the photosphere \citealp{Duvall1979,Lo2010,See2016,Vidotto2018}.
To bring the model closer to the realistic cases we consider the harmonic
magnetic field approximation for the outer magnetic field \citep{Bonanno2016},
i.e., 
\begin{equation}
\left(\nabla^{2}+k^{2}\right)\overline{\boldsymbol{B}}=0,\label{eq:harm}
\end{equation}
for the external region $r_{e}<r<2.5R$, and the radial magnetic field
for $r\ge2.5R$. We assume that $\left(kR\right)^{2}\ll$1. At the
surface, we employ the continuity of the normal component of the magnetic
field and the tangential component of the mean electromotive force.
For the azimuthal component of the vector potential outside the dynamo
domain, we seek a solution in the form of the decomposition of a product
of the spherical Bessel functions and associated Legendre polynomials
as follows (cf. \cite{Bonanno2016}), 
\begin{equation}
A\left(x,\theta,t\right)=\sum A^{(n)}\left(t\right)\frac{\left(\gamma^{(n)}j_{n}\left(x\xi\right)+y_{n}\left(x\xi\right)\right)}{\left(\gamma^{(n)}j_{n}\left(x_{e}\xi\right)+y_{n}\left(x_{e}\xi\right)\right)}\sin\theta P_{n}^{1}\left(\theta\right),\label{eq:Aex}
\end{equation}
where $x=r/R$, $\xi=kR$ and $x_{e}=0.99$ is the external boundary
of the dynamo domain; the constants $A^{(n)}\left(t\right)$ and $\gamma^{(n)}$
are determined by the condition of continuity of the radial magnetic
field at $x_{e}$: 
\begin{equation}
\frac{\partial A}{\partial x}=\sum\sin\theta P_{n}^{1}\left(\theta\right)A^{(n)}\left(t\right)\left(\frac{n}{x_{e}}-\xi\frac{\left(\gamma^{(n)}j_{n+1}\left(x_{e}\xi\right)+y_{n+1}\left(x_{e}\xi\right)\right)}{\left(\gamma^{(n)}j_{n}\left(x_{e}\xi\right)+y_{n}\left(x_{e}\xi\right)\right)}\right),\label{eq:Ad}
\end{equation}
and the outer coronal magnetic field boundary condition, for instance,
the pure radial magnetic field at the radius of the source surface.
We place this point at $x_{s}=2.5$, where we define $\gamma^{(n)}$:
\[
\frac{n}{x_{s}}-\xi\frac{\left(\gamma^{(n)}j_{n+1}\left(x_{s}\xi\right)+y_{n+1}\left(x_{s}\xi\right)\right)}{\left(\gamma^{(n)}j_{n}\left(x_{e}\xi\right)+y_{n}\left(x_{e}\xi\right)\right)}=0.
\]
The continuity of the tangential component of the mean electromotive
force determines the magnitude of the toroidal magnetic field at the
surface. For the axisymmetric toroidal component, the external magnetic
field decomposition is similar to Eq(\ref{eq:Aex}): 
\begin{equation}
B\left(x,\theta,t\right)=\sum B^{(n)}\left(t\right)\frac{\left(\zeta^{(n)}j_{n}\left(x\xi\right)+y_{n}\left(x\xi\right)\right)}{\left(\zeta^{(n)}j_{n}\left(x_{e}\xi\right)+y_{n}\left(x_{e}\xi\right)\right)}P_{n}^{1}\left(\theta\right),\label{eq:Bex}
\end{equation}
where $\zeta^{(n)}$ is deduced from the condition at $x_{s}$, $B\left(x_{s},\theta,t\right)=0$.
At the top of the dynamo domain, we require the continuity of $\left[\mathcal{E}_{\theta}\right]_{x=x_{e}}=0$
and the same for the toroidal magnetic field. This results in the
following boundary condition: 
\begin{equation}
\eta_{T}\frac{\partial xB}{\partial x}\!=\!\eta_{T}^{(+)}\sum P_{n}^{1}\left(\theta\right)B^{(n)}\!\!\left(\!n+1\!-\!\xi x_{e}\frac{\left(\zeta^{(n)}j_{n+1}\left(x_{e}\xi\right)+y_{n+1}\left(x_{e}\xi\right)\right)}{\left(\zeta^{(n)}j_{n}\left(x_{e}\xi\right)+y_{n}\left(x_{e}\xi\right)\right)}\!\right),\label{eq:Bd}
\end{equation}
where $\eta_{T}^{(+)}$ is the effective turbulent diffusion in the
corona surrounding the dynamo domain. For the case $\eta_{T}^{+}\gg\eta_{T}$
and $\xi,k=0$ , we return to the case of the vacuum boundary conditions.
In what follows, we study the effect of the ratio $\eta_{T}^{+}/\eta_{T}$
and magnitude of the parameter $\xi$ on the dynamo instability and
helical properties of the coronal magnetic field. 

To study the dynamo instability, we represent the dynamo solution in
form $A\sim\hat{A}\exp\left(\lambda t\right)$ and employ $B\sim\hat{B}\exp\left(\lambda t\right)$
where the eigen value $\lambda$ s is complex; its real part determines
the instability of the dynamo mode and the imaginary part defines the
oscillatory parameters. The eigenfunctions $\hat{A}$ and $\hat{B}$
are complex, as well. They are define the spatial structure of the
eigen dynamo modes. The integration in radius and latitude is done
with the help of the Galerkin method. In the radial direction we decompose
$\hat{A}$ and $\hat{B}$ into Chebyshev polynomials using the Gauss-Lobatto
grid with 50 mesh points, and in the latitudinal direction we employ
the associated Legendre polynomials $P_{n}^{1}\left(\theta\right)$
and the Gauss-Legendre grid with 72 points from pole to pole. To satisfy
the radial boundary conditions, we use the basis recombination method
\citep{Boyd2001} . The code is written in \textsc{python} employing
the linear algebra libraries of the \textsc{python} \textsc{numpy}
and \textsc{scipy} packages \citep{2020SciPyNMeth}. 

\section{Results}

\subsection{The dynamo instability}

\begin{figure*}
\includegraphics[width=0.9\textwidth]{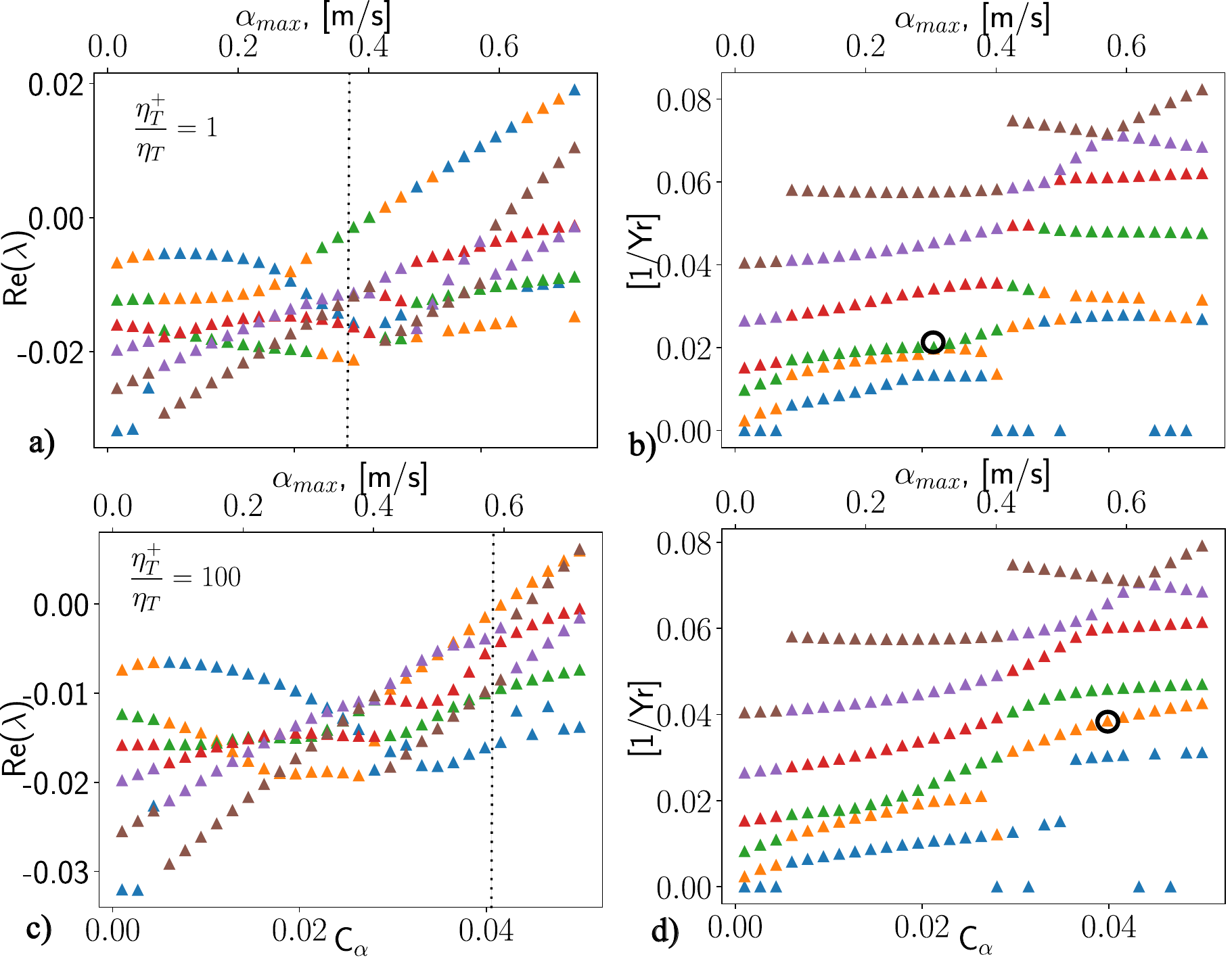}

\caption{a) Growth rates of the first six eigen odd dynamo modes for the solar
type dynamo model for the ratio $\eta_{T}^{+}/\eta_{T}=1$, the x-axis
show the maximum magnitude of the $\alpha_{\phi\phi}$ component in
the convection zone; colors mark the different eigen modes; b) shows
the eigen frequency for each dynamo mode, the circles mark the first
unstable modes; c) and d) show the same as a) and b) for ratio $\eta_{T}^{+}/\eta_{T}=100$.}\label{fig:local}
\end{figure*}

As a first step, we consider the eigenvalue problem. It helps us
to define the critical thresholds of the dynamo instability and the
eigenmode profiles. Figure \ref{fig:local} shows the growth rates
and frequencies for the first six eigenmodes for the dynamo model.
The instability diagram shows the modes of the odd parity, i.e., those
that are antisymmetric about the equator. The even modes show very
similar diagrams, but they have a higher instability threshold than
the odd modes. Also, we did not find them in the non-linear runs. Therefore, we omit their discussion below. We see that for the ratio $\eta_{T}^{+}/\eta_{T}=1$
the instability threshold is about 1.5 lower than for the case $\eta_{T}^{+}/\eta_{T}=100$.
The latter show the critical magnitude of the $\alpha$-effect which is
close to the case of the vacuum boundary condition. The dynamo instability diagram for the vacuum boundary condition was considered by \cite{Pipin23MN},
see Figure 2 there. We see that in the case $\eta_{T}^{+}/\eta_{T}=1$
the dynamo period is about 40 years and the case $\eta_{T}^{+}/\eta_{T}=100$
shows twice as less the dynamo period for the first unstable mode.

\begin{figure}
\includegraphics[width=0.9\textwidth]{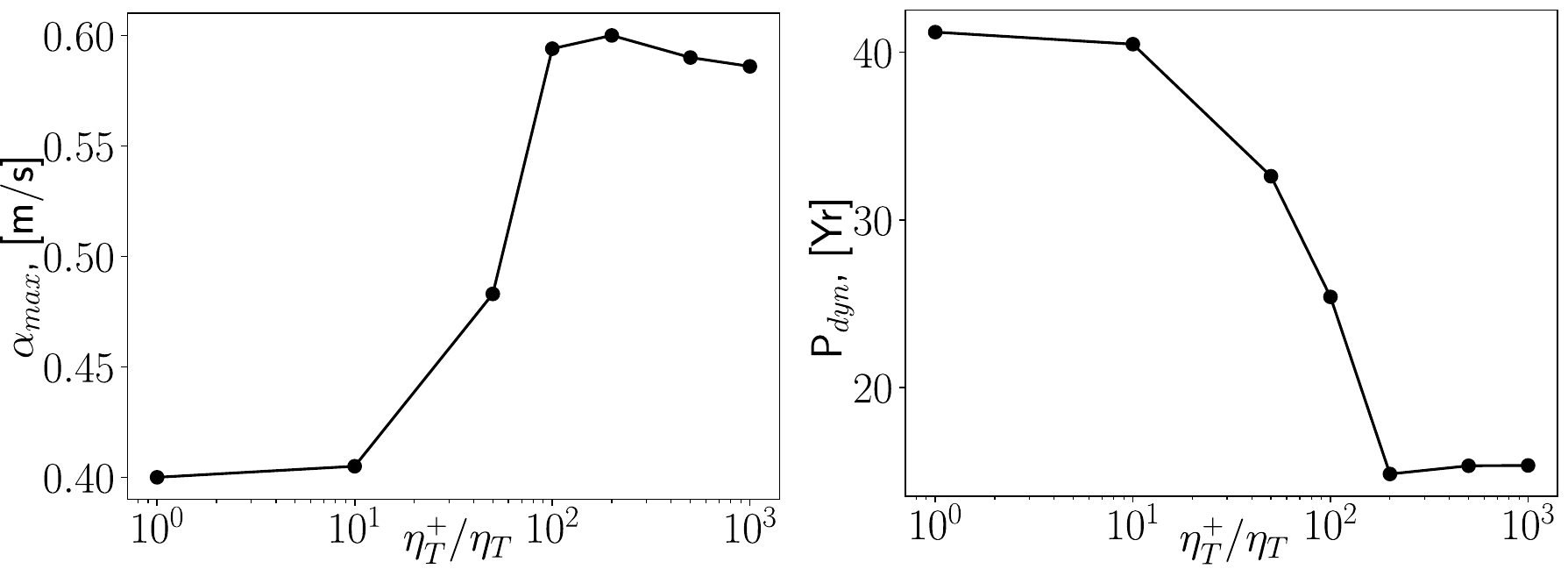}

\caption{a) Variation of the critical amplitude of the $\alpha$ effect in
the convective convective envelope of the Sun, depending on the ratio
$\eta_{T}^{+}/\eta_{T}$; b) the same as a) for variation of the dynamo
period.}\label{fig:Fig3}

\end{figure}

Figure \ref{fig:Fig3} shows the variations of the dynamo instability
threshold and the dynamo period of the first unstable mode for the
range of ratio $\eta_{T}^{+}/\eta_{T}$ between 1 and 1000. For
the high ratio $\eta_{T}^{+}/\eta_{T}\gg1$, the instability threshold
tends to the limit of the vacuum boundary conditions where the dimensionless $\alpha$-effect parameter $C_{\alpha}\approx0.04$, which corresponds
to the maximum amplitude of the $\alpha_{max}\approx0.6$ m/s in the
solar convection zone. The dynamo period decreases to the value of about
14 years. 
%It should be noted that the eigenvalue problem is solved for the dynamo model which is restricted by the boundary of the convective envelope. 

We have tried to consider the low values of $\eta_{T}^{+}/\eta_{T}<1$.
Contrary to \cite{Elstner2020IAUS} we find no changes for the
critical threshold and the dynamo period in decrease $\eta_{T}^{+}/\eta_{T}$
below 1. This shows that coronal differential rotation, which is included
in their model, has a strong impact on the results of the dynamo runs.
In nonlinear runs, we exclude $\eta_{T}^{+}/\eta_{T}<1$ from our study.

\subsection{The structure and helicity of dynamo waves}

\begin{figure}
\includegraphics[width=0.8\textwidth]{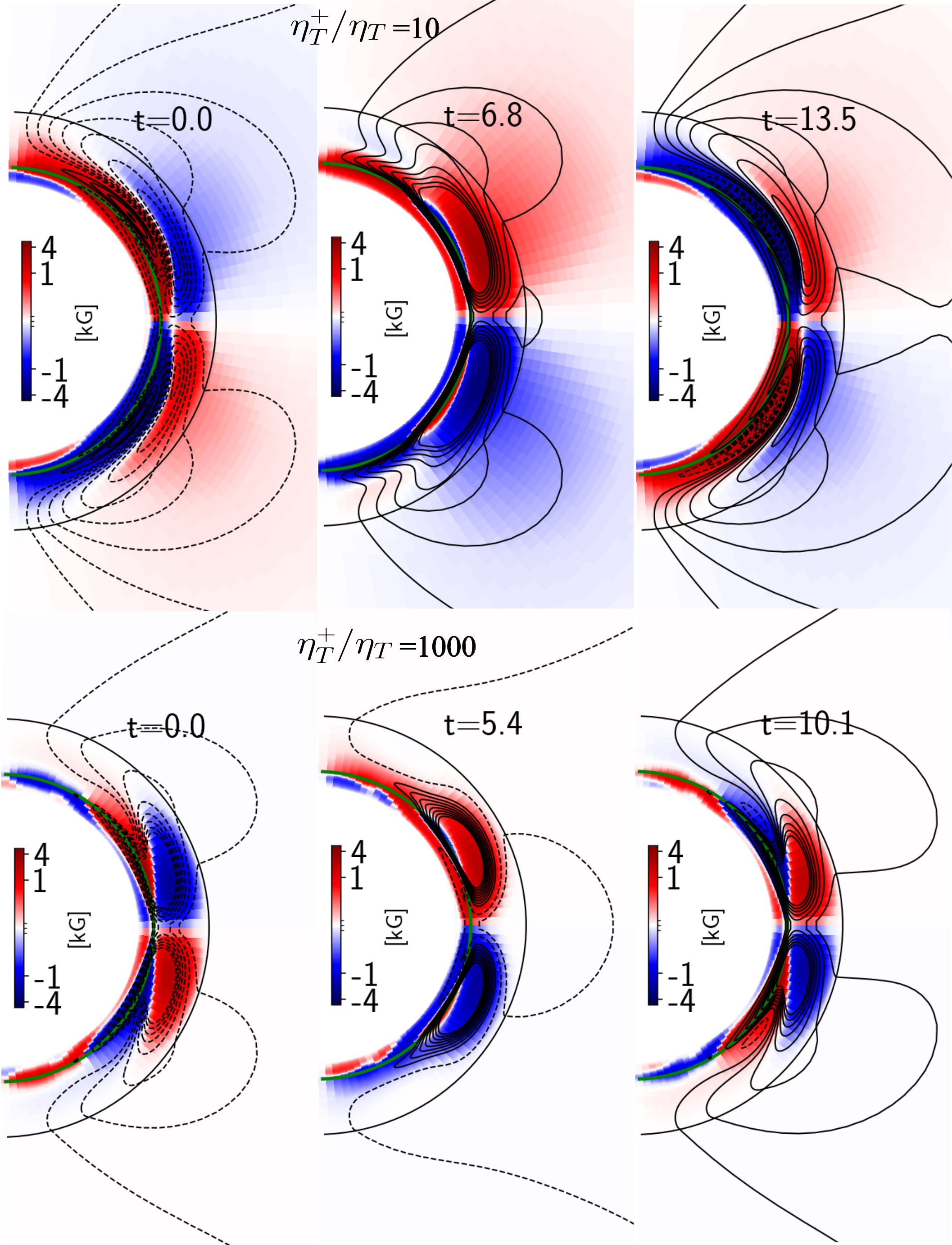}

\caption{Snapshots of the magnetic field distributions inside and outside
of the solar convection zone for the half magnetic cycle: top row
show the case $\eta_{T}^{+}/\eta_{T}=10$ and the bottom shows the
same for $\eta_{T}^{+}/\eta_{T}=1000$. The color image shows the
toroidal magnetic field; to reflect the weak toroidal magnetic field
in the solar corona we employ the logarithmic density plot with the
linear threshold of 100 G; the contours show streamlines of the poloidal
magnetic field; solid lines show the clockwise streamlines; the green
line show the bottom of the convection zone.}\label{fig:Snaps}

\end{figure}

In the rest of the paper we discuss the properties of the nonlinear
dynamo runs under change of the parameter $\eta_{T}^{+}/\eta_{T}$.
The nonlinear dynamo model includes the overshoot layer with distribution
of $\alpha$-effect and turbulent diffusivity parameters as shown
in Figure \ref{fig1}. In all runs, we use the same amplitude of the
$\alpha$ effect and put it above the critical level of case $\eta_{T}^{+}/\eta_{T}=1000$, ie $C_{\alpha}=0.042$. The run starts from a weak poloidal magnetic field
that consists of a mixture of dipole and quadrupole parity. The antisymmetric about
the equator magnetic field establishes after about 10 magnetic cycles.
Figure \ref{fig:Snaps} shows the snapshots of the dynamo waves for
the half magnetic cycle in two cases: $\eta_{T}^{+}/\eta_{T}=10$
and $\eta_{T}^{+}/\eta_{T}=1000$. We see that in the first case the
dynamo wave of the toroidal magnetic field can extend over the whole
range of latitudes from the equator to poles. In the second case,
$\eta_{T}^{+}/\eta_{T}=1000$, we see that dynamo activity can
have up to three waves over the hemisphere. 

\begin{figure}
\includegraphics[width=0.99\textwidth]{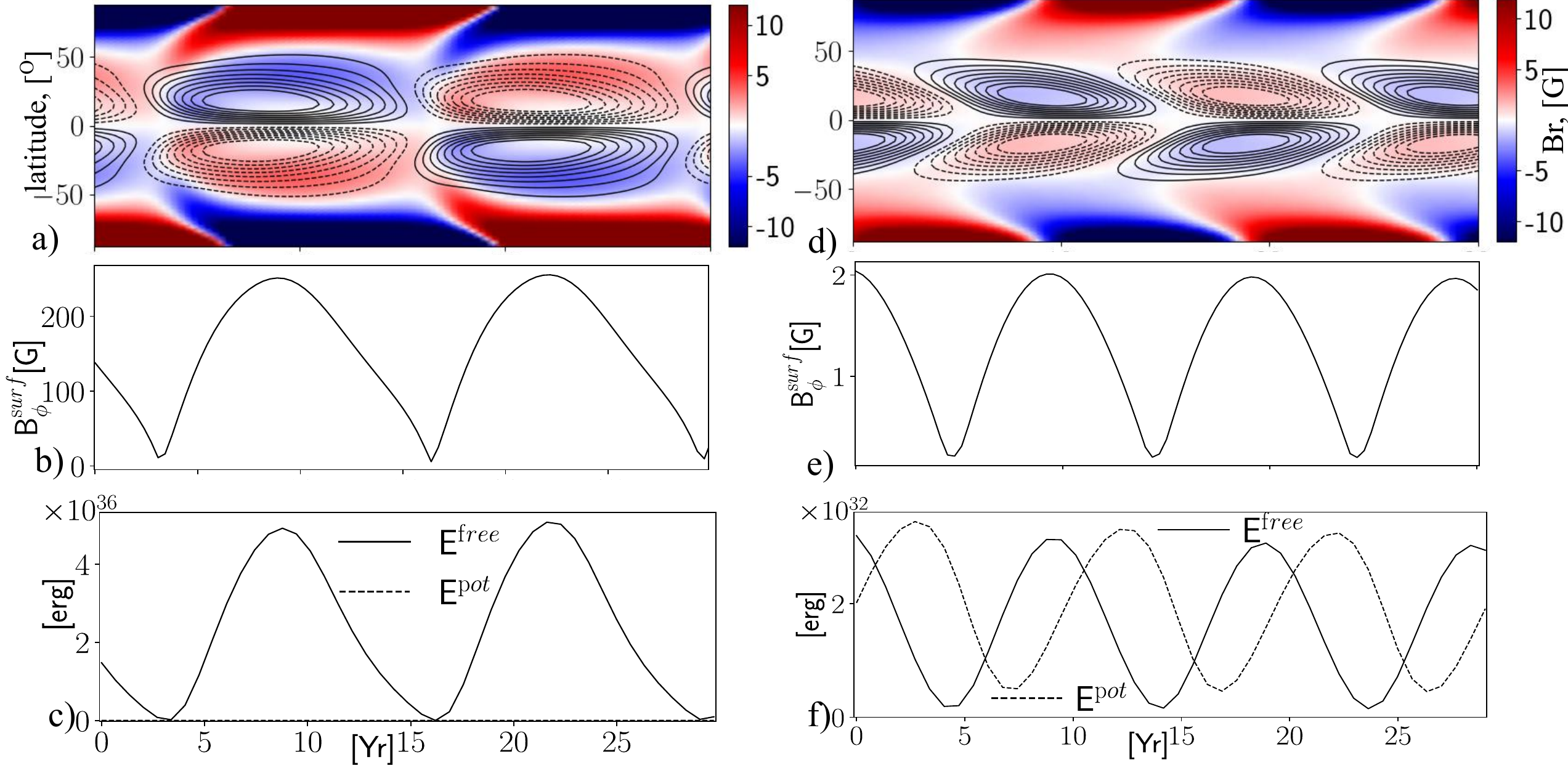}

\caption{a) The time-latitude diagram of the surface radial magnetic field
magnetic helicity density (color image) and contours in range of $\pm1$kG
show the the near-surface toroidal magnetic field at $r=0.9R$ for
ratio$\eta_{T}^{+}/\eta_{T}=10$ ; b) evolution of the mean surface
toroidal magnetic field; c) evolution of the total energy of the potential
magnetic fields, $E^{pot}$, dashed line, and the total free magnetic
energy in the stellar corona, $E^{free}$ , solid line ; panels d),
e) and f) show the same as a), b) and c) for the ratio$\eta_{T}^{+}/\eta_{T}=1000$. }\label{fig:bfl}

\end{figure}

Figures \ref{fig:bfl} a) and d) show the time-latitude diagrams for the
near-surface toroidal magnetic field at $r=0.9R$ and the surface
radial magnetic field in the runs with the ratio parameter $\eta_{T}^{+}/\eta_{T}=10$
and $\eta_{T}^{+}/\eta_{T}=1000$, respectively. For the low ratio
$\eta_{T}^{+}/\eta_{T}$, we find a nearly steady wave pattern over
latitudes. In fact, the previous Figure\ref{fig:Snaps} indicates the
radial propagation of the dynamo waves in this case. The dynamo period
in this case is about 30 years. This run shows the magnitude of the
surface toroidal magnetic field of about 200 G, see Figure\ref{fig:bfl}
b). The history of the magnetic observations does not show evidence
for such a strong toroidal magnetic on the solar photosphere. The
model run for the high ratio $\eta_{T}^{+}/\eta_{T}=1000$ shows the
standard solar-like variations of the toroidal magnetic field, which
reflects sunspot activity at the low latitudes and the two-branches
activity of the radial magnetic field: the polar branch with a period
of 20 years and the same long-term extended wave for the equatorial
branch. Both are in good agreement with the analysis observations
of \cite{2005AA438349K,Pipin2021b}. The magnitude of the surface toroidal
magnetic field in this model agrees with the results of \cite{Lo2010,Vidotto2018}.

Using the results of the coronal magnetic field simulation, we estimate
the total free magnetic energy in the solar corona.
\begin{eqnarray}
E^{free} & = & E^{tot}-E^{pot},\label{eq:free}
\end{eqnarray}
where
\begin{eqnarray}
E^{tot} & = & \frac{1}{8\pi}\int\boldsymbol{\overline{B}}^{2}\mathrm{d}V,\label{eq:tot}\\
E^{pot} & = & \frac{1}{8\pi}\int\boldsymbol{\overline{B}}^{(pot)2}\mathrm{d}V,\label{eq:pot}
\end{eqnarray}
where integration is done over the bulk of corona in range $r=.99R-2.5R$
, $\boldsymbol{\overline{B}}^{(pot)}$ stands for the potential part
of the coronal magnetic field. It is determined using the potential
field extrapolation between the surface radial magnetic field and
the source surface at $r=2.5R$ , where the coronal magnetic field
becomes pure radial. Figures \ref{fig:bfl} c) and f) indicate that the
parameter $E^{free}$ remains intact with the near-surface toroidal magnetic field. In the run with $\eta_{T}^{+}/\eta_{T}=10$ the $E^{free}/E^{pot}\sim10^{4}$
and $E^{pot}\sim10^{32}$erg. The latter magnitude is consistent with the
estimations of \cite{Yeates2024SoPh} for solar Cycle 24.
The results of the run with $\eta_{T}^{+}/\eta_{T}=1000$ show $E^{free}/E^{pot}\sim1$.
In this case, the value $E^{free}\sim10^{32}$erg is about an order of
magnitude less than that found in the paper cited above. It should
be noted that calculations of \cite{Yeates2024SoPh} disregard
the mean surface toroidal magnetic field. In that model the main source
of the free magnetic energy in the corona is due to the solar
active regions activity, which we neglect in our study. 

\begin{figure}
\includegraphics[width=0.9\textwidth]{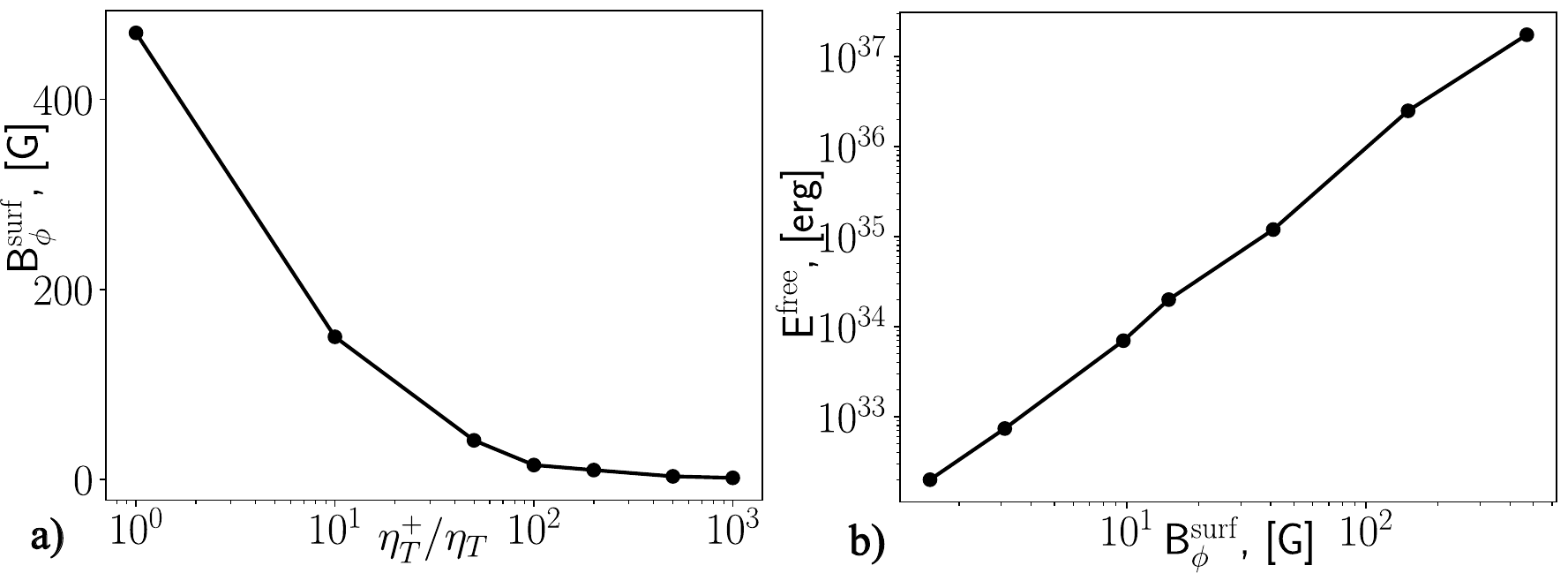}

\caption{a) Dependence of the amplitude of the toroidal magnetic field on
the surface, on the ratio $\eta_{T}^{+}/\eta_{T}$j; b) Dependence
of the free energy of the corona, $E^{free}$ ,on amplitude of the
surface toroidal magnetic field, $B_{\phi}^{surf}$.}\label{fig:dep}

\end{figure}

We calculate a series on the nonlinear dynamo runs to determine the dependence
of the amplitude of the surface toroidal magnetic field, $B_{\phi}^{surf}$
, on the ratio $\eta_{T}^{+}/\eta_{T}$. Figure \ref{fig:dep} a)
indicates that $B_{\phi}^{surf}$ decreases to zero with an increase
of $\eta_{T}^{+}/\eta_{T}$. In other words, the limit of the current-free
boundary conditions is fulfilled in the nonlinear runs as well. The
amplitude of $E^{free}$ shows the power-law dependence on the strength
of the surface magnetic field,
\begin{equation}
E^{free}\sim0.3R^{3}\left|B_{\phi}^{surf}\right|^{2}.\label{eq:pow}
\end{equation}

\begin{figure}

\includegraphics[width=0.9\textwidth]{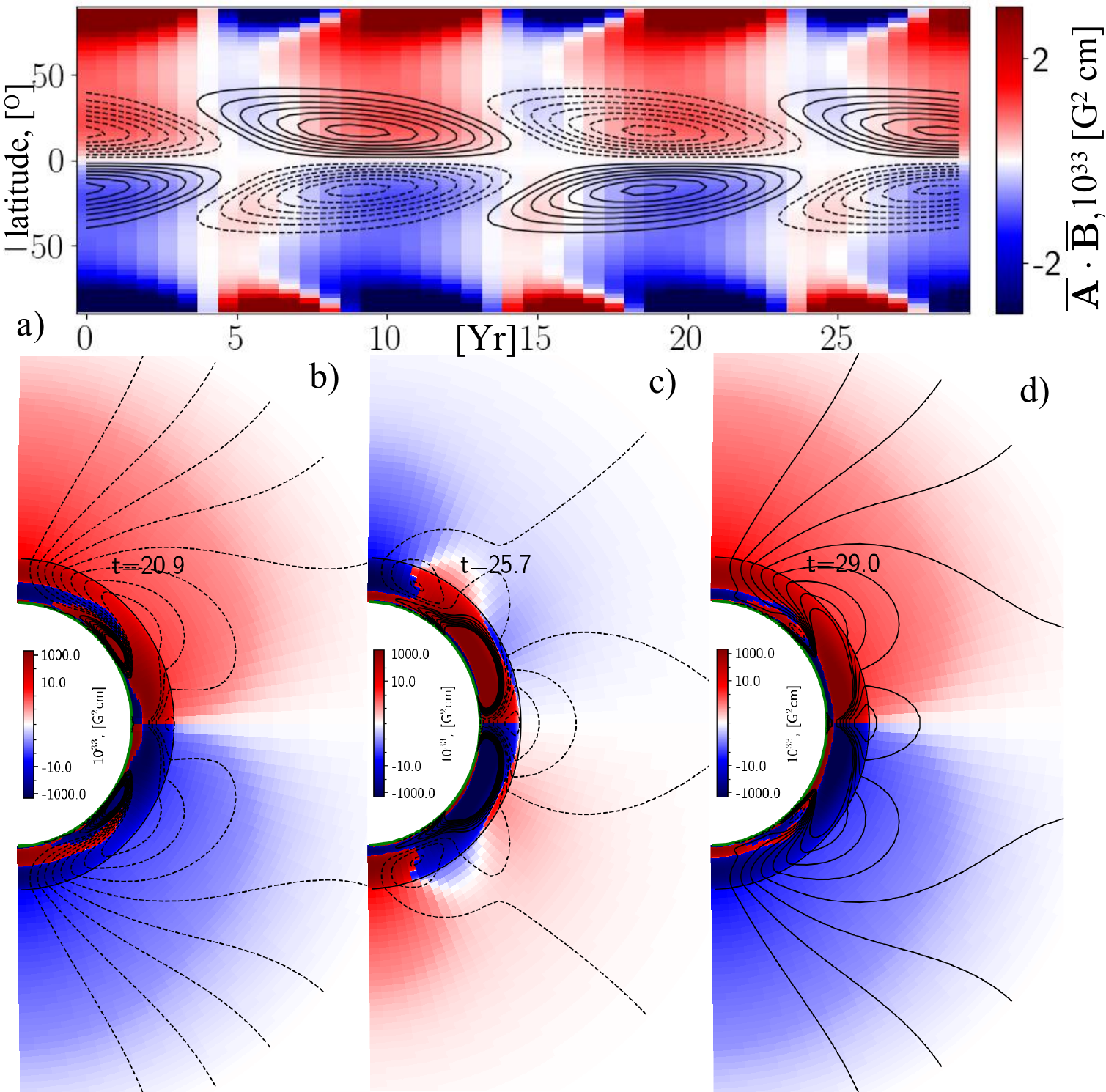}\caption{a) The time - latitude variations of the surface magnetic helicity
density (color image) and the near surface toroidal magnetic field
at $r=0.9R$ (contours in range of $\pm1$kG); the bottom row shows
snapshots of the magnetic helicity density in the solar convective
zone and corona for the half magnetic cycle. }\label{fig:mh}

\end{figure}
Finally, Figure \ref{fig:mh} shows evolution of the large-scale magnetic
field helicity density at the stellar surface and in overall volume
including the convection zone and corona. The magnetic helicity density
pattern satisfies the helicity rule, showing the positive helicity
in the northern hemisphere and the opposite sign of helicity in the southern hemisphere.
The helicity of the coronal magnetic field follows that rule. The
helicity rule inverses during the beginning of each magnetic cycle.
This effect is also supported by solar observations \citep{Pevtsov2021}.
The model shows that this inversion of the magnetic helicity affects
the variation of the magnetic helicity sign along the radius. For
example, the snapshot Figure \ref{fig:mh} c) shows the positive helicity
density at the northern mid-latitudes at the solar surface and the
negative one in the upper stellar corona. Analysis of the Ulysses
observations reveals the change of the helicity sign with radius as
well \citep{Brandenburg2011}. In addition, they found the predominantly
positive magnetic helicity of the small-scale magnetic field in the
solar wind. Their analysis covers the period of the solar Cycle 22
minimum. The model shows that for this period of time the magnetic helicity of the solar corona can have a negative sign at the north. Following the balance of the total magnetic helicity, we can speculate that the
small-scale magnetic helicity should have the opposite sign to the
helicity of the large-scale magnetic field. This conclusion needs
a support from a better dynamo model which consistently takes the stellar
wind into account.

\section{Discussion and conclusions}

In our paper, we discuss the distributed dynamo model, which can be
one of the preferable choices among solar dynamo models \citep{Brandenburg2005,BRetal23}.
The present results reveal for the first time that the structure and
period of the dynamo waves inside the convection zone can depend on
the turbulent diffusivity contrast between the stellar convection
zone and stellar corona. As expected by \cite{Bonanno2016}, this
result holds for the distributed stellar dynamos. In addition, we demonstrate
a theoretical possibility for dynamical coupling between the dynamo
inside the convective envelope and the active processes in the stellar
corona by means of variations in the diffusive properties of the stellar
corona. For the harmonic boundary conditions of the magnetic field, such a coupling can be parameterized by the ratio $\eta_{T}^{+}/\eta_{T}$.
Our model presents a simple alternative for the dynamo models of \cite{JakBran2021AA,Perri2021ApJ}
and has to be further justified by using more realistic coronal
models following the lines of the above-cited papers. 

The dynamo models show an increase in the surface toroidal magnetic
in a decrease of $\eta_{T}^{+}/\eta_{T}$. For the case of  $\eta_{T}^{+}/\eta_{T}=1$,
the surface toroidal field reaches $0.5$ kG. Such a strong toroidal
field is observed in fast-rotating solar-type stars \citep{See2016}.
It can provide a lot of free energy for coronal magnetic activity.
We see that for this case the free energy exceeds $10^{34}$erg, which
is the typical energy of super-flares in solar analogs with a faster
rotation rate than the Sun \citep{Okamoto2021ApJ}. For the Sun,
the ratio $\eta_{T}^{+}/\eta_{T}=1000$ fits the weak, $\sim1-2$ G toroidal magnetic field on the surface. Whether and how much the mean ratio $\eta_{T}^{+}/\eta_{T}$ can change in time should be further investigated.

Another effect of the harmonic boundary condition is that it provides
the magnetic helicity flux from the dynamo region to the corona. The
model shows that the coronal magnetic field helicity varies nearly
intact with the helicity of the large-scale magnetic field on the
surface. The preferable hemispheric sign of the large-scale magnetic
helicity is determined by the $\alpha$-effect \citep{Blackman2003}.
It is mainly positive in the northern hemisphere of the solar convection
zone; see Figure\ref{fig1}. However, the negative helicity of the
large-scale magnetic field is expected for the periods of the magnetic-cycle minima. This is due to the different phases of the magnetic field
evolution in the mid-latitudes and in the polar regions of the star.
This results in the opposite helicity sign between the surface and
the coronal magnetic field for this period. It should be noted that
the given mechanism of the large-scale magnetic helicity production
is specific to the mean-field dynamo distributed over the solar convection
zone. The Bawcock-Leighton dynamo model operates with the nonlocal
$\alpha$-effect \citep{Hazra2023a}. The helicity evolution of
the large-scale magnetic field in this case is different from that shown
in Figure \ref{fig:mh}. Moreover, the effects of the activity of the bipolar active regions can substantially modify our picture\citep{PShKo2025}.

In summary, the application of the harmonic boundary condition shows
that variations of the mean diffusive properties of the stellar corona
can provide strong feedback on the large-scale dynamo inside the
convective envelope. The solar corona is probably close to the ideal
dielectric state with an effective diffusivity jump $\eta_{T}^{+}/\eta_{T}\ge1000$
which is reflected in a weak toroidal magnetic field at the photosphere.
The surface variations of the large-scale magnetic helicity in the
distributed convective zone $\alpha^{2}\Omega$ dynamo can result
in the radial inversion of the magnetic helicity in the stellar corona.

\begin{acknowledgements}
The author thanks the Ministry of Science and Higher Education of
the Russian Federation for financial support (Subsidy No.075-GZ/C3569/278)
\end{acknowledgements}
 
\begin{authorcontribution}
The author takes full responsibility for results  presented in the paper.
\end{authorcontribution}

\begin{fundinginformation}
The author thanks the Ministry of Science and Higher Education of
the Russian Federation for financial support (Subsidy No.075-GZ/C3569/278)
\end{fundinginformation}

\begin{dataavailability}
The data of the dynamo simulations are available from the author  by
request.
\end{dataavailability}

\begin{materialsavailability}
All additional materials related to the paper can be available from
the author by request.
\end{materialsavailability}

\begin{codeavailability}
We put the eigenvalue code, which is written in python, in the \textsc{zenode} archive. Using the code, the reader can reproduce the results presented in Figures\ref{fig:local} and \ref{fig:Fig3}The dynamo code is available by request from the author.
\end{codeavailability}

\begin{ethics}
\begin{conflict}
The author declares that he has no conflicts of interest.
\end{conflict}
\end{ethics}
\bibliographystyle{spr-mp-sola}
\bibliography{dyn}

\end{document}